\documentclass[usegraphicx]{mn2e}

\usepackage[fleqn]{amsmath}
\usepackage{amssymb}
\def\d{\mathrm{d}}
\newcommand{\plm}{\ensuremath{\pm\ }}
\def\kms{\mathrm{\,km\, s^{-1}}}
\def\kmsM{\mathrm{\,km\, s^{-1}\, Mpc^{-1}}}

\title[FP and $M/L$ evolution lens galaxies]{The Fundamental Plane and
the evolution of the $M/L$ ratio of early-type field galaxies up to
$z \sim 1$}  
\author[van de Ven, van Dokkum \& Franx]{
  G.~van~de~Ven,$^1$\thanks{E-mail: glenn@strw.leidenuniv.nl}\ 
  P.G.~van~Dokkum,$^2$\thanks{Present address: Department of
    Astronomy, Yale University, P.O. Box 208101, New Haven, CT 
    06520-8101}\  
  M.~Franx$^1$\\
  $^1$Sterrewacht Leiden, Postbus 9513, 2300 RA Leiden, The
  Netherlands \\ 
  $^2$California Institute of Technology, MS105-24, Pasadena, CA
  91125}  
\date{Accepted 0000 Month 00. Received 0000 Month 00;in original 0000 Month 00}
\pagerange{\pageref{firstpage}--\pageref{lastpage}}
\pubyear{0000}

\begin{document}

\label{firstpage}

\maketitle


\begin{abstract}

We analyse the Fundamental Plane (FP) of 26 strong gravitational lens
galaxies with redshifts up to $z \sim 1$, using tabulated data from
Kochanek et al.\ (2000\nocite{2000ApJ...543..131K}) and
Rusin et al.\ (2003\nocite{2003ApJ...astroph0211229}).
The lens galaxies effectively form a mass-selected sample of early-type
galaxies in environments of relatively low density.
We analyse the FP and its evolution in the restframe Johnson $B$ and
Gunn $r$ bands.  
Assuming that early-type galaxies are a homologous family, the FP
then provides a direct measurement of the $M/L$ ratio evolution. 

If we assume that the $M/L$ ratios of field early-type galaxies evolve
as power-laws, we find for the lens galaxies an evolution rate 
$\d\log(M/L)/\d z$ of $-0.62 \pm 0.13$ in restframe $B$ 
and $-0.47 \pm 0.11$ in restframe Gunn $r$ 
for a flat cosmology with $\Omega_M=0.3$ and $\Omega_\Lambda=0.7$. 
For a Salpeter (1955\nocite{1955ApJ...121..161S}) Initial Mass Function
and Solar metallicity these results correspond to mean stellar
formation redshifts of 
$\langle z_\star \rangle= 1.8_{-0.5}^{+1.4}$ and 
$1.9_{-0.6}^{+1.9}$ respectively.
After correction for maximum progenitor bias,
van Dokkum \& Franx (2001\nocite{2001ApJ...553...90V}) 
find a mean stellar formation redshift for cluster galaxies
of  $\langle z_\star^{cl} \rangle = 2.0_{-0.2}^{+0.3}$, 
which is not significantly different from that found for the lens
galaxies.  
However, if we impose the constraint that lens and cluster galaxies
that are of the same age have equal $M/L$ ratios and we do not correct
for progenitor bias, the difference is significant and we find that
the stellar populations of the lens galaxies are 10--15$\,\%$ younger
than those of the cluster galaxies.

We find that both the $M/L$ ratios as well as the restframe colors of
the lens galaxies show significant scatter. 
About half of the lens galaxies are consistent with an old
cluster-like stellar populations, but the other galaxies are bluer and
best fit by single burst models with younger stellar formation
redshifts as low as $z_\star \sim 1$. 
Moreover, the scatter in color is correlated with the scatter in $M/L$
ratio.
We interpret this as evidence of a significant age spread among the
stellar populations of lens galaxies, whereas the ages of the stellar
populations of the cluster galaxies are well approximated by a single
formation epoch.

\end{abstract}



\section{Introduction}
\label{sec:introduction}

One of the central problems in astronomy is that of galaxy
formation and evolution: when were the visible parts of galaxies
assembled and when were the stars formed? The merging of galaxies
leads to changes in their masses, and stellar evolution
changes their luminosities. The evolution of the
mass-to-light ($M/L$) ratio relates the mass and luminosity
evolution. 

Galaxy mass measurements are notoriously difficult.
Fortunately, empirical relations such as the Tully-Fisher relation
for spiral galaxies 
(Tully \& Fisher 1977\nocite{1977A&A....54..661T}) 
and the Fundamental Plane (FP) for early-type galaxies 
(Dressler et al.\ 1987\nocite{1987ApJ...313...42D}; 
Djorgovski \& Davis 1987\nocite{1987ApJ...313...59D}) 
can provide us with information on the masses and mass evolution of
galaxies. 
The FP is a tight relation between the structural parameters and
velocity dispersion, which under the assumption of homology implies
that $M/L \propto M^\frac{1}{4}$ with low scatter (Faber et
al. 1987\nocite{1987Faber}).   

Due to stellar evolution the $M/L$ ratio of a stellar population
changes with redshift, and hence also the FP will change.
The redshift evolution of the intercept of the FP is proportional to
the evolution of the mean $M/L$ ratio. 
Hence, the tightness of the FP relation makes it a very sensitive
indicator of the mean age of the stellar population of early-type
galaxies (van Dokkum \& Franx 1996\nocite{1996MNRAS.281..985V}). 

For cluster galaxies, the $M/L$ ratio evolves very slowly, indicating
that the stars were formed at redshifts $z\gtrsim 3$
(e.g. Kelson et al.\ 1997\nocite{1997ApJ...478L..13K}; 
Bender et al.\ 1998\nocite{1998ApJ...493..529B}; 
van Dokkum et al.\ 1998\nocite{1998ApJ...504L..17V}).   
Current semi-analytical hierarchical models place the assembly
time of typical early-type galaxies at much lower redshifts
(e.g. Kauffmann 1996\nocite{1996MNRAS.281..487K}; Kauffmann \& Charlot
1998\nocite{1998MNRAS.294..705K}; Diaferio et
al. 2001\nocite{2001MNRAS.323..999D}).  
However, in hierarchical galaxy formation models the assembly time and
star formation epoch are strongly dependent on the environment, with
cluster early-type galaxies forming much earlier than those in the
general field (e.g. Kauffmann 1996). 
Hence cluster early-type galaxies do not provide the most stringent
tests of these models.
Moreover, the measured evolution of early-type galaxies may
underestimate the true evolution because of the effects of
morphological evolution. 
If many early-type galaxies evolved from late-type galaxies at
$z\lesssim 1$, the sample of early-type galaxies at high redshift is
only a subset of all progenitors of present-day early-type galaxies.
This would lead us to underestimate the luminosity evolution, and
hence overestimate the stellar formation redshift
(see van Dokkum \& Franx 2001\nocite{2001ApJ...553...90V}, 
hereafter vDF01).  

Recent studies have started to explore the FP and $M/L$ ratios of high
redshift early-type galaxies in the general field.  
Interestingly, the FP of field early-type galaxies appears to be quite
similar to that of cluster early-type galaxies out to $z\approx0.55$ 
(Treu et al.\ 2001\nocite{2001MNRAS.326..237T}; 
van Dokkum et al.\ 2001\nocite{2001ApJ...553L..39V}), 
in apparent conflict with current semi-analytical models (see van
Dokkum et al.\ 2001\nocite{2001ApJ...553L..39V}). 
However, there are indications for a significant offset between the two
populations at higher redshift 
(Treu et al.\ 2002\nocite{2002ApJ...564L..13T}).
  
Kochanek et al.\ (2000\nocite{2000ApJ...543..131K}, hereafter K00)
analysed the FP and color evolution of strong gravitational lens
galaxies up to $z\sim1$.  
The lensing cross section is dominated by galaxies with high central velocity
dispersions, and the lens galaxies effectively provide a mass-selected
sample of field early-type galaxies which can be compared to optically
selected samples of field and cluster galaxies. 
The mass-selection is important because it limits Malmquist-type
biases, and it is much less sensitive to selection effects caused by
morphological evolution. 
K00 find that the stars constituting the lens galaxies must have
formed at $z \gtrsim 2$ for a flat cosmology with $\Omega_M=0.3$ and 
$\Omega_{\Lambda}=0.7$, and conclude from their analysis that there
are no significant differences between field and cluster early-type
galaxies. 

K00 analyse the evolution of the FP in the observed photometric 
bands, and their modeling necessarily includes the large variation in
observed magnitudes and colors due to redshift (the ``K''-correction). 
This complicates the measurement of the smaller effect due to
evolution of (the stellar populations of) the lens galaxies.
Rusin et al.\ (2003\nocite{2003ApJ...astroph0211229}, hereafter R03)
study the same sample of lens galaxies, with recent photometric
observations included.
They convert the data from observed filters into magnitudes in
standard restframe bands, so that the evolution of the
$M/L$ ratio that follows from the FP (under the usual assumption that
early-type galaxies are a homologous family) can be investigated
instead of the evolution within the FP. 
This clarifies the analysis and allows for direct comparison of the
results with those of other FP studies.  
From their $M/L$ evolution analysis R03 find a (2$\sigma$) lower limit
$z > 1.8$ for the formation of the stars in lens galaxies.
Although this limit is more precise, it is similar to that of
K00. Hence, R03 also conclude that the evolution measurements favor
old stellar populations among field galaxies, like those of cluster
galaxies, and argue against significant episodes of star formation at
$z<1$, as predicted by the semi-analytical hierarchical models. 

For the analysis in this paper we use the tabulated data of K00,
extended with the recent photometric measurements as given by R03.
We convert the observed surface brightnesses and colors to
the restframe Johnson $B$ and Gunn $r$ bands.
Our transformation to restframe differs from that of R03.
In both approaches the modeled color between restframe band and
observed filter is used to convert the observed magnitude into an
estimate of the restframe magnitude.
R03 obtain the modeled color for a given spectral energy distribution,
whereas we use the observed color between a pair of filters to  
interpolate between the modeled colors for four different spectral
types (Fig. \ref{fig:Magtrafo}).
Moreover, while R03 use the (weighted) contribution of all observed
magnitudes to determine the restframe magnitude, we select the best
one with small observational error and filter close to the
(redshifted) restframe band.
Hence, the model dependence of our correction is small and we exclude
observed magnitudes with large uncertainties, minimizing the error in
the resulting restframe magnitude.
We analyze the $M/L$ evolution of the lens galaxies and compare
our results with those of R03 and with results from previous studies.  
Additionally, we test whether age differences between the lens galaxies
are significant. 
To this end, we study the scatter in both the $M/L$ evolution and the
restframe colors of the lens galaxies, and investigate whether the
deviations are correlated.

This paper is organized as follows. 
In Section \ref{sec:FPpar} we determine the FP parameters, using the
lensing geometry to estimate the velocity dispersion.   
The transformation from observed to restframe bands is described in
Section \ref{sec:trafo2restrame}.  
In Section \ref{sec:fpandmlevo}, we construct the FP of the lens
galaxies and present the $M/L$ evolution derived from the FP. 
We estimate the age of the stellar populations of the lens galaxies by
fitting single burst models to the $M/L$ evolution in Section
\ref{sec:ages}.  
In Section \ref{sec:colors} we study the colors of the lens galaxies.  
The results are summarized and discussed in Section
\ref{sec:discconcl}. 
Unless stated otherwise we assume $H_0=50\kmsM$ and a
flat cosmology with $\Omega_M=0.3$ and $\Omega_\Lambda=0.7$. 
We note that our results are not dependent on the value of the Hubble
constant.


\section{FP parameters}  
\label{sec:FPpar}

\begin{table*}
  \begin{center}
    \begin{tabular}{lcccccc}
      \hline
      Name Lens    & $z$ & $\sigma_c$ & $\sigma_{c\star}$ & $r_e$ & $\mu_{e,B_z}$ & $\mu_{e,r_z}$ \\
      & & ($\mathrm{km\,s^{-1}}$)   &  ($\mathrm{km\,s^{-1}}$) & (kpc) & 
      ($\mathrm{mag\,arcsec^{-2}}$) & ($\mathrm{mag\,arcsec^{-2}}$) \\
      \hline
      0047-2808     & 0.49 & 254 \plm 26 & 229 \plm 15 &  7.7 \plm 0.7 & 21.79 \plm 0.16 & 21.11 \plm 0.28 \\ 
      Q0142-100     & 0.49 & 224 \plm 22 &             &  4.3 \plm 0.2 & 20.42 \plm 0.05 & 19.64 \plm 0.05 \\ 
      MG0414+0534   & 0.96 & 303 \plm 30 &             &  8.6 \plm 1.6 & 21.07 \plm 0.17 & 19.97 \plm 0.16 \\ 
      B0712+472     & 0.41 & 181 \plm 18 &             &  2.8 \plm 0.4 & 20.67 \plm 0.17 & 19.60 \plm 0.16 \\ 
      RXJ0911+0551  & 0.77 & 260 \plm 26 &             &  7.0 \plm 0.6 & 21.74 \plm 0.17 & 20.75 \plm 0.14 \\ 
      FBQ0951+2635  & \textit{0.24} & 128 \plm 13 &             &  0.9 \plm 0.2 & 19.82 \plm 0.25 & 19.05 \plm 0.25 \\ 
      BRI0952-0115  & \textit{0.41} & 117 \plm 12 &             &  0.8 \plm 0.2 & 20.00 \plm 0.21 & 18.61 \plm 0.20 \\ 
      Q0957+561     & 0.36 & 431 \plm 43 & 305 \plm 11 & 14.1 \plm 1.3 & 22.48 \plm 0.12 & 21.45 \plm 0.11 \\ 
      LBQS1009-0252 & \textit{0.88} & 198 \plm 21 &             &  1.9 \plm 0.3 & 19.81 \plm 0.14 & 18.90 \plm 0.13 \\ 
      Q1017-207     & \textit{0.78} & 151 \plm 16 &             &  3.1 \plm 0.1 & 21.28 \plm 0.49 & 20.28 \plm 0.07 \\ 
      FSC10214+4724 & \textit{0.75} & 241 \plm 26 &             & 11.8 \plm 5.2 & 22.97 \plm 0.44 & 22.03 \plm 0.56 \\ 
      B1030+074     & 0.60 & 218 \plm 22 &             &  4.2 \plm 0.6 & 21.38 \plm 0.14 & 20.14 \plm 0.25 \\ 
      HE1104-1805   & 0.73 & 316 \plm 32 &             &  6.4 \plm 1.9 & 21.34 \plm 0.33 & 20.22 \plm 0.30 \\ 
      PG1115+080    & 0.31 & 210 \plm 21 & 288 \plm 27 &  3.0 \plm 0.1 & 21.23 \plm 0.06 & 20.01 \plm 0.05 \\ 
      HST14113+5211 & 0.46 & 190 \plm 19 &             &  3.8 \plm 0.4 & 21.70 \plm 0.09 & 20.72 \plm 0.11 \\ 
      HST14176+5226 & 0.81 & 292 \plm 29 & 230 \plm 14 &  7.5 \plm 0.9 & 20.98 \plm 0.15 & 20.35 \plm 0.12 \\ 
      B1422+231     & 0.34 & 160 \plm 16 &             &  2.1 \plm 0.6 & 21.18 \plm 0.25 & 19.93 \plm 0.24 \\ 
      SBS1520+530   & 0.72 & 220 \plm 22 &             &  3.5 \plm 0.3 & 20.13 \plm 0.17 & 19.36 \plm 0.08 \\ 
      MG1549+3047   & 0.11 & 188 \plm 19 & 242 \plm 20 &  2.3 \plm 0.2 & 21.39 \plm 0.09 & 20.12 \plm 0.09 \\ 
      B1608+656     & 0.63 & 292 \plm 29 &             &  6.2 \plm 1.0 & 20.68 \plm 0.19 & 19.88 \plm 0.15 \\ 
      MG1654+1346   & 0.25 & 206 \plm 21 &             &  4.9 \plm 0.1 & 21.99 \plm 0.07 & 20.73 \plm 0.05 \\ 
      MG2016+112    & 1.00 & 299 \plm 30 & 328 \plm 32 &  2.5 \plm 0.3 & 19.22 \plm 0.10 & 18.06 \plm 0.10 \\
      B2045+265     & 0.87 & 378 \plm 38 &             &  4.1 \plm 1.3 & 20.50 \plm 0.44 & 19.41 \plm 0.30 \\
      HE2149-2745   & 0.50 & 203 \plm 20 &             &  4.3 \plm 0.4 & 20.92 \plm 0.12 & 20.36 \plm 0.12 \\
      Q2237+030     & 0.04 & 168 \plm 17 & 220 \plm 31 &  4.3 \plm 0.8 & 22.37 \plm 0.50 & 21.15 \plm 0.23 \\
      HS0818+1227   & 0.39 & 251 \plm 25 &             &  6.6 \plm 0.2 & 22.18 \plm 0.07 & 21.13 \plm 0.04 \\
      \hline                                           
   \end{tabular}                      
   \caption[]{\slshape                
     FP parameters of 26 strong gravitational lens galaxies with
     redshifts up to $z\sim1$. 
     For 5 lens galaxies the redshift is not known spectroscopically,
     and a photometrically estimated value (in italics) is given. 
     The velocity dispersion $\sigma_c$, within the standard aperture
     with a diameter of $3\farcs4$ at the distance of the Coma
     cluster, follows from the lensing geometry, assuming a singular
     isothermal sphere mass model and a 
     Hernquist (1990\nocite{1990ApJ...356..359H}) luminosity profile.  
     The velocity dispersion from stellar kinematics $\sigma_\star$
     has been measured for 7 lens galaxies (references in text).   
     The effective radius $r_e$ and effective surface brightness
     $\mu_e$ follow from fits to an $r^{1/4}$ law.  
     To allow a direct comparison with the local FP, the effective
     surface brightness has been corrected to the restframe Johnson
     $B$ and Gunn $r$ band by interpolating between filters. 
   }
   \label{tab:fppar}
  \end{center}
\end{table*}

The study of the FP of strong gravitational lens galaxies differs in
two important aspects from that of cluster galaxies. 
First, the lens galaxies are individual galaxies spread over a large
range in redshifts, instead of an ensemble of galaxies at the same 
redshift. 
Additionally, whereas studies of the FP of optically selected galaxies
measure velocity dispersions from spectra, for lens galaxies we use
the lensing geometry to estimate this quantity.
The separation between lensed images of background sources increases
with the mass of the lens and is therefore a measure of the velocity
dispersion.  
The two remaining FP parameters, the effective radius and surface
brightness, are determined from surface photometry as for cluster
galaxies.

\subsection{Velocity dispersion}
\label{sec:veldisp}

For a singular isothermal sphere (SIS) mass model the relation between
the velocity dispersion of the matter distribution and the separation
of the source images $\Delta \theta$ is 
$\Delta\theta = 8 \pi (\sigma_D/c)^2
D_{\mathrm{LS}}/D_{\mathrm{OS}}$. 
Here, $D_{\mathrm{LS}}$ and $D_{\mathrm{OS}}$ are the angular
diameter distances from the lens galaxy to the source and
from the observer to the source. 
To determine these values, the redshifts of both the lens galaxy and
the source are needed 
(e.g. Hogg 2000\nocite{2000...astroph9905116v4}). 
If the redshift of a lens galaxy is not known spectroscopically, we
adopt a photometrically estimated value\footnote{Obtained from the
  CfA-Arizona Space Telescope Lens Survey (CASTLES) web site at
  http://cfa-www.harvard.edu/castles/} 
with a $10\,\%$ uncertainty. 
The lens systems for which no redshift is known for the source,
we exclude from our analysis.  
This leaves a total of 26 lens galaxies, of which 5 have a redshift
that is estimated photometrically (Table \ref{tab:fppar}). 
The velocity dispersion $\sigma_D$ depends also (weakly) on
cosmology through the angular diameter distances, but is independent
of the value of the Hubble constant since the distances appear as a
ratio. 

The velocity dispersion $\sigma_D$ is that of the total matter
distribution, including possible dark matter, rather than the central
velocity dispersion of the stellar component $\sigma_c$. 
For a given mass and luminosity distribution, the ratio $g \equiv
\sigma_c/\sigma_D$ can be modeled by solving the single 
Jeans equation for a spherical system with $\sigma_\phi=\sigma_\theta$
and $\beta=1-\sigma_\theta^2/\sigma_r^2$ the anisotropy parameter 
(e.g. Binney \& Tremaine 1987\nocite{1987gady.book.....B}).
The overall mass distribution is assumed to be isothermal. 
We model the luminosity distribution by a 
Hernquist (1990\nocite{1990ApJ...356..359H}) 
profile with characteristic radius $a=r_e/1.8153$. 
For each lens galaxy with effective radius $r_e$ in kpc 
(see Section \ref{sec:effradandsurfbright}), we integrate the
resulting line of sight velocity dispersion within the Coma aperture
to obtain an estimate for $g$, so that the stellar velocity dispersion
follows as $\sigma_c=g\,\sigma_D$.

For the sample of 26 lens galaxies, $\sigma_D$ is distributed with a
(biweight\footnote{Throughout this paper we use the biweight location and scale  
as estimators of the mean and standard deviation (rms) respectively. 
These estimators are robust for a broad range of non-Gaussian
underlying populations and are less sensitive to outliers than
standard estimators (e.g. Beers, Flynn \& Gebhardt
1990\nocite{1990AJ....100...32B}).}) 
mean of $230\kms$ and a (biweight) standard deviation of $61\kms$.
In the case of an isotropic system ($\beta=0$), the estimated stellar
velocity dispersion $\sigma_c$ has a mean of $223\kms$ and a standard
deviation of $70\kms$.
Hence, the properties of the sample of lens galaxies will be typical
for early-type field galaxies close to $L*$, which have a
characteristic velocity dispersion around $225\kms$ 
(e.g. Kochanek 1994\nocite{1994ApJ...436...56K},
1996\nocite{1996ApJ...466..638K}).  

Several sources contribute to the uncertainty in $\sigma_c$.
We account for the scatter in the image separation $\Delta\theta$,
which is small and set to 2\%. 
Also the error in the effective radius $r_e$ (Table \ref{tab:fppar})
contributes through the modeled ratio $g$. 
For those galaxies for which only a photometric estimate of the lens
redshift $z_l$ is available, we included an error of $10\,\%$. 
Since the angular diameter distance depends on redshift this error is
an additional contribution to the uncertainty in $\sigma_c$ via the
determination of $\sigma_D$ and via the conversion of the effective
radius in arcseconds into physical units of kpc.   
Apart from these errors in observational parameters, we also want to
take into account that we have made several assumptions in the
modeling of the ratio $g$.
For the 26 lens galaxies, the distribution of $g$ has a mean of $0.90$,
$0.94$ and $1.01$ for an anisotropy parameter $\beta$ of $-0.5$, $0$
and $0.5$ respectively.  
For each lens galaxy we apply the isotropic case $\beta=0$, but to
take the variation of $g$ with $\beta$ into account, together with the
assumption of a SIS mass model and a Hernquist profile, we assume an
additional error of $10\,\%$ in $g$ and hence in $\sigma_c$.   

For 7 out of the 26 lens galaxies presented in this paper, a velocity
dispersion measured from stellar kinematics $\sigma_{c\star}$ is
available: 
0047-2808     (Koopmans \& Treu 2003\nocite{2003ApJ...583..606K}),
Q0957+561     (Tonry \& Franx 1999\nocite{1999ApJ...515..512T}; 
               Falco et al.\ 1997\nocite{1997ApJ...484...70F};
               Rhee 1991\nocite{1991Natur.350..211R}),
PG1115+080    (Tonry 1998\nocite{1998AJ....115....1T}), 
HST14176+5226 (Ohyama et al.\ 2002\nocite{2002AJ....123.2903O}),
MG1549+3047   (Leh\'ar et al.\ 1996\nocite{1996AJ....111.1812L}),
MG2016+112    (Koopmans \& Treu 2002\nocite{2002ApJ...568L...5K}) and  
Q2237+030     (Foltz et al.\ 1992\nocite{1992ApJ...386L..43F}).
Comparing the measured velocity dispersions (Table \ref{tab:fppar})
with the modeled $\sigma_c$ from the lensing geometry, we find for
these 7 galaxies that the ratio $\sigma_{c\star}/\sigma_c$ is
distributed with a mean of $1.07$ and dispersion of $0.27$.
For the sample of 7 lens galaxies both methods are consistent.
However, due to peculiarities of the lensing system, such as the
contamination of Q0957+561 by the mass distribution of the underlying
cluster, the velocity dispersion from both methods can be
significantly different for individual lens galaxies.

\subsection{Effective radius and surface brightness}
\label{sec:effradandsurfbright}

The lensing systems were observed with the \texttt{WFPC2},
\texttt{NICMOS1} and \texttt{NICMOS2} camera on the \texttt{HST}, 
in filters ranging from the visual F555W through the infrared F205W
filter. 
For each system, the image with optimal contrast between the lens
galaxy and the images of the background source is selected, and the
effective radius $r_e$ and the mean surface brightness within
the effective radius $\langle\mu_e\rangle$ from fits to an $r^{1/4}$ law is
determined. 
The data and model fits are described in detail by
Leh\'ar et al.\ 2000\nocite{2000ApJ...536..584L} and K00.
Note that $r_e$ depends slightly on cosmology due to the conversion of
the effective radius from units of arcseconds from the fit into physical
units of kpc.  
Moreover, since the lens redshift is used in the conversion, we have
to take into account a (small) additional contribution to the
uncertainty in $r_e$ due to the assumed $10\,\%$ error in the lens
redshift in the case it is estimated photometrically 
(see also Section \ref{sec:veldisp}).  

In this paper we use the surface brightness at the effective radius
$\mu_e$ (in $\mathrm{mag\,arcsec^{-2}}$), which is related to
$\langle\mu_e\rangle$ by $\mu_e-\langle \mu_e \rangle = 1.393$.  
We also define $I_e \equiv 10^{-\mu_e/2.5}$.
In the following, we refer to the filter in which the fit was made as
the reference filter. 
The dependence of effective radius on passband due to color gradients
can be ignored because of  the strong correlation between $\mu_e$ and
$r_e$ (see Section \ref{sec:fpandmlevo}), and we can use the observed
colors (tabulated by K00) to calculate the effective surface
brightness in each filter.  

The effective surface brightnesses and colors are corrected for Galactic
extinction with an $R_V \equiv A(V)/E(B-V) \approx 3.1$ extinction
curve for a diffuse stellar medium 
(e.g. Cardelli, Clayton \& Mathis 1989\nocite{1989ApJ...345..245C}; 
O'Donnell 1994\nocite{1994ApJ...422..158O}). 
The galactic extinction $E(B-V)$ is obtained from 
Schlegel, Finkbeiner \& Davis (1998\nocite{1998ApJ...500..525S}).
Extinction corrections on the effective surface brightness are
typically $\sim 0.03$ magnitudes in the reference filter, which
in most cases is the F160W filter.
The three galaxies MG0414+0534, MG2016+112 and B2045+265 are
exceptions with significantly higher galactic extinction of 0.18, 0.14
and 0.14 magnitudes respectively.


\section{Transformation to restframe}
\label{sec:trafo2restrame}

In order to compare the FP of the redshifted lens galaxies directly
to the FP of the Coma cluster at $z=0.023$, we calculate the
effective surface brightness of the lens galaxies in restframe 
Johnson $B$ band and restframe Gunn $r$ band by interpolating between
filters. 

In the following example we assume that the redshifted $r$ band falls
between the \texttt{WFPC2} F555W (=$V$) and F814W (=$I$) filters and
the effective surface brightness is determined in the \texttt{NICMOS}
F160W (=$H$) reference filter. 
We assume a linear relation between the AB magnitudes of the
redshifted $r$ filter and the $V$ and $I$ filter, so that for the
restframe $r$ magnitude of a galaxy at redshift $z$ we can write 
\begin{equation}
  \label{eq:rz}
  r_z = V - \alpha(V-I) - \alpha(c_V-c_I) + c_V-c_r + 2.5\log(1+z).
\end{equation}
The constants $c$ are the conversion constants between the standard
Vega magnitudes and AB magnitudes
\begin{equation}
  \label{eq:convconst}
  c = 2.5\log\left( \frac{\int_0^\infty T(\nu)
  f_\nu^{\mathrm{Vega}}\d\nu }{ \int_0^\infty T(\nu) \d\nu} \right)
  -48.60, 
\end{equation}
with $f_\nu^{\mathrm{Vega}}$ the fluxdensity of Vega, and $T(\nu)$ the
filter transmission. 
The transmission curves of the \texttt{HST} filters (including the CCD
response) were obtained from \textsc{STScI}\footnote{
  http://www.stsci.edu/instruments/observatory/cdbs/cdbs.html }, 
and those of the $B$ and $r$ passbands were obtained from Bessel
(1990\nocite{1990PASP..102.1181B}) and Thuan \& Gunn
(1976\nocite{1976PASP...88..543T}) respectively.       
We used the \texttt{CALCPHOT} task of the \texttt{STSDAS} package in
\texttt{IRAF} to calculate the conversion constants for the
\texttt{HST} filters. 
The conversion constants for the $B$ and $r$ passband follow from
Bessel (1990\nocite{1990PASP..102.1181B}) and Frei \& Gunn
(1994\nocite{1994AJ....108.1476F}) respectively.  
The last term in \eqref{eq:rz} includes the broadening of the $r$ band
with redshift and makes the magnitude behave as if it is a flux,
rather than a fluxdensity. 

\begin{figure}
  \begin{minipage}[t]{4cm}
    \begin{center}
      \includegraphics[draft=false,scale=0.22,trim=0cm 3cm 0cm 1cm]{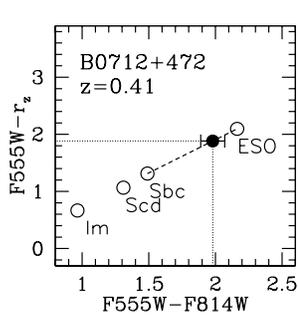} 
      \label{fig:Magtrafo}
    \end{center}
  \end{minipage}
  \begin{minipage}[t]{4cm}
  \caption[]{\slshape 
    Transformation to restframe $r$ band by interpolation between the
    F555W (=$V$) and F814W (=$I$) filter. For four spectral types the
    $V-I$ and $V-r_z$ colors are modeled (open circles). With the
    observed $V-I$ color of the lens galaxy (filled circle) we
    interpolate between the modeled colors of the two nearest spectral
    types, to find the $V-r_z$ color of the lens galaxy. 
    }
  \end{minipage}
\end{figure}

Equation \eqref{eq:rz} relates the $V-r_z$ color of a lens galaxy to
its observed $V-I$ color. 
To determine $\alpha$, we model the $V-I$ and $V-r_z$ colors for
four different spectral types, E/S0, Sbc, Scd and Im, using the
spectral energy distribution of Coleman, Wu \& Weedman
(1980\nocite{1980ApJS...43..393C}).  
This gives four estimates of $\alpha$. 
Using the observed $V-I$ color of the lens galaxy we interpolate
linearly between the modeled colors of the two nearest spectral
types (Fig. \ref{fig:Magtrafo}). 
In this way we obtain the best-fit $V-r_z$ color and corresponding
value of $\alpha$ for the lens galaxy.  
To estimate the uncertainty in the conversion constants
we compare modeled colors with those predicted by
Frei \& Gunn (1994\nocite{1994AJ....108.1476F}). 
We find that the differences are small and estimate the uncertainty at 0.02
magnitudes.   

The observed $V-H$ color is then used to relate the effective surface
brightness in the reference $H$ filter to that of the restframe $r$
band   
\begin{equation}
  \label{eq:murz}
  \mu_{e,r_z} = \mu_{e,H} + (V-H) - (V-r_z).  
\end{equation}
Similar transformations are derived for each lensing system.
In cases where observations are available in more than two passbands, we
calculate $\mu_{e,r_z}$ for all filter combinations and make a
selection based on the following criteria: the error in $\mu_{e,r_z}$, 
the wavelength difference between the redshifted $r$ band and the
observed filter, and a preference for filter pairs enclosing the
redshifted $r$ band. 
The latter implies an interpolation between two filters, whereas
otherwise we have to extrapolate. 
As byproducts of our procedure we find the SEDs that provide the best
fits to the observed colors of the lens galaxies. 
For the restframe $B$ and $r$ band this yields respectively 20 
($77\,\%$) and 19 ($73\,\%$) lens galaxies that are best fitted by the
E/S0 spectral type, whereas the colors of the remaining lens galaxies
are closest to the Sbc spectral type.    

Since $r_z$ in \eqref{eq:rz} behaves like a flux, the effective
surface brightness decreases as $(1+z)^4$ with
increasing redshift\footnote{The restframe $r_z$ magnitude
  \eqref{eq:rz} behaves therefore as a $K$-corrected magnitude.}. 
We correct the surface brightnesses for this cosmological dimming.
For the 26 lens galaxies, the resulting values for the restframe
Johnson $B$ and Gunn $r$ effective surface brightness $\mu_{e,B_z}$
and $\mu_{e,r_z}$ are given in Table \ref{tab:fppar}, together with
the other two FP parameters; the central stellar velocity dispersion
$\sigma_c$ and effective radius $r_e$.  
We have also included the redshift $z$ of the lens galaxies and the
velocity dispersions measured from stellar kinematics
$\sigma_{c\star}$ when available.


\section{FP and $M/L$ Evolution}
\label{sec:fpandmlevo}

\begin{figure}
  \begin{center}
    \includegraphics[draft=false,scale=0.45,trim=0cm 1.5cm 0cm 2.5cm]{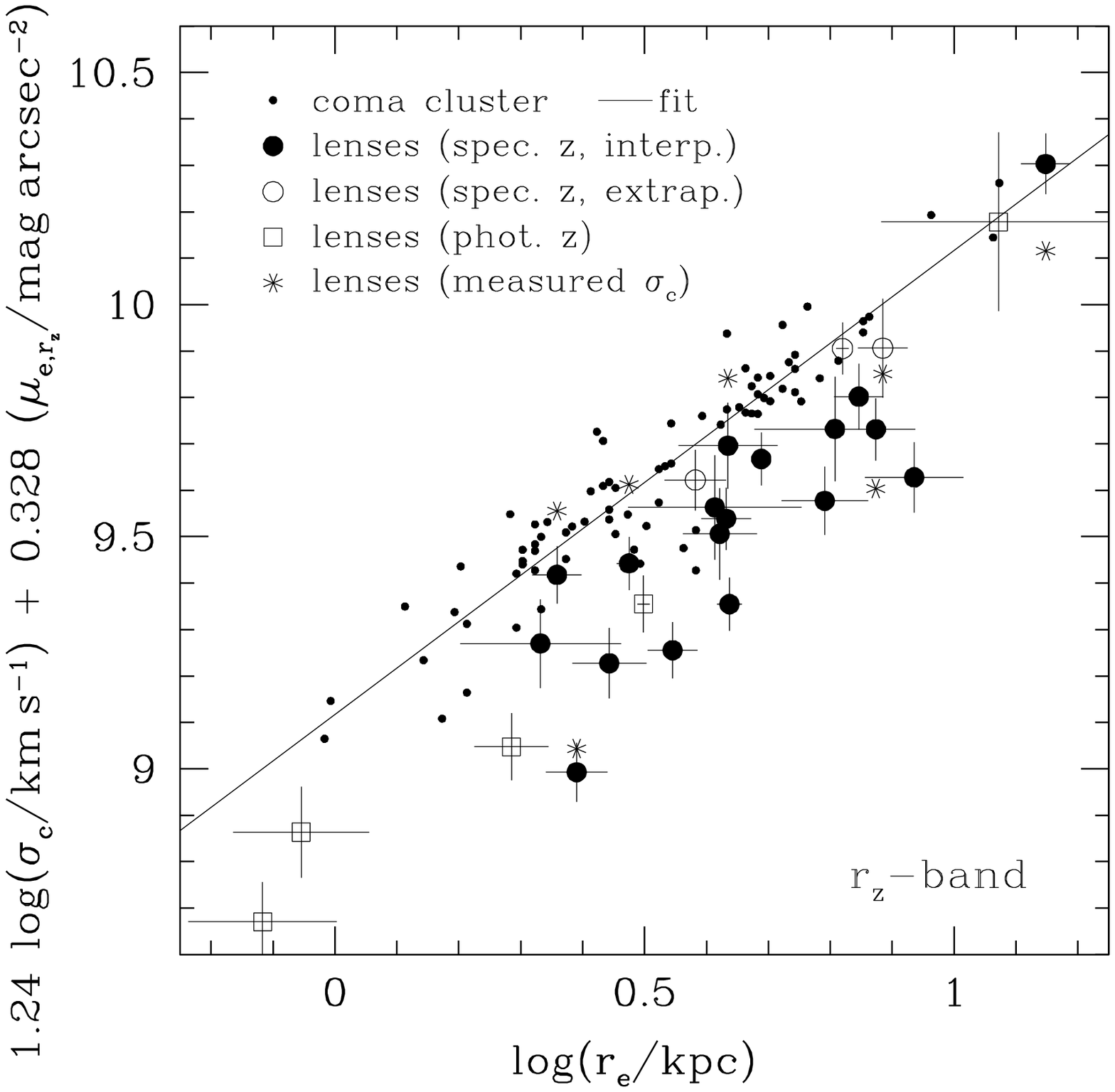} 
  \end{center}
  \caption[]{\slshape 
    Edge-on view of the FP in the restframe Gunn $r$ band.
    The Coma cluster galaxies (small dots) and corresponding
    linear fit represent the local FP 
    (J\o rgensen, Franx and Kj\ae rgaard 1995a\nocite{1995MNRAS.273.1097J},
    1995b\nocite{1995MNRAS.276.1341J}, 1996\nocite{1996MNRAS.280..167J}).
    Lens galaxies for which the redshifts are known spectroscopically
    are indicated with circles.
    Filled circles correspond to lens galaxies for which the transformation
    to restframe (Section \ref{sec:trafo2restrame}) was by interpolation
    between a pair of filters, and open circles if a modest
    extrapolation was needed.   
    The open squares represent the 5 lens galaxies with a redshift
    that is estimated photometrically (Table \ref{tab:fppar}).
    The transformation to restframe for these 5 galaxies was by
    interpolation. 
    For 7 lens galaxies, velocity dispersions have been
    measured from stellar kinematics (see text for details).
    Asterisks show the position of these galaxies in the FP if the stellar
    velocity dispersion is used.
    The lens galaxies are offset from the Coma FP, as expected from
    evolution of their stellar populations.
  }   
  \label{fig:FPr}
\end{figure}

The Fundamental Plane has the form
\begin{equation}
  \label{eq:fprelB}
  \log r_e = \alpha\log\sigma_c + \beta\log I_e + \gamma
\end{equation}
(Dressler et al.\ 1987\nocite{1987ApJ...313...42D}; 
Djorgovski \& Davis 1987\nocite{1987ApJ...313...59D}), with
$r_e$ in kpc, $\sigma_c$ in $\kms$ and $\mu_e$ ($=-2.5\log I_e$) in
$\mathrm{mag\,arcsec^{-2}}$.  
We adopt for the coefficients $\alpha$ and $\beta$ the values derived
by  J\o rgensen et al.\ (1996\nocite{1996MNRAS.280..167J}) 
for a sample of 225 early-type galaxies in nearby clusters. 
They found for the Johnson $B$ band $\alpha=1.20\pm0.06$ and
$\beta=-0.83\pm0.02$, and for the Gunn $r$ band $\alpha=1.24\pm0.07$
and $\beta=-0.82\pm0.02$. 

We use the tabulated photometric and spectroscopic data of 
J\o rgensen et al.\ 
(1995a\nocite{1995MNRAS.273.1097J}, 1995b\nocite{1995MNRAS.276.1341J})
to construct the FP of Coma. 
The edge-on projection of the Coma FP in the $r$ band is shown in
Fig. \ref{fig:FPr} (small dots).
A linear fit to the coma FP yields an intercept of $9.50\pm0.02$ and
$9.12\pm0.02$ for the $B$ and $r$ band respectively.
Large symbols show the lens galaxies. 
The lens galaxies show a large scatter, and are offset with respect to
the FP of Coma.
This relative difference can be attributed to the evolution of the
$M/L$ ratios of galaxies, and the large scatter may in part be caused
by the large range of redshifts spanned by the lens sample. 

As usual in FP evolution studies, we assume that early-type galaxies
are a homologous family, i.e. that they are structurally similar.
The total mass of a galaxy (including possible dark matter) is then
proportional to an effective mass $\propto \sigma_c^2 r_e$. 
With the total luminosity proportional to $I_e r_e^2$, the FP
relation implies that the effective mass-to-light ratio 
$M/L \propto M^{0.24} r_{e}^{-0.02}$ in the $r$ band 
(e.g. Treu et al.\ 2001\nocite{2001MNRAS.326..237T}).
The tightness of the FP relation implies a low scatter in the $M/L$
ratios of early-type galaxies of $23\,\%$  
(Faber et al.\ 1987\nocite{1987Faber}; J\o rgensen, Franx \& Kj\ae rgaard
1996\nocite{1996MNRAS.280..167J}).
Hence, the evolution of $M/L$ can be well studied via the evolution of
the FP.

\begin{figure*}
  \includegraphics[draft=false,scale=1.0,trim=0.5cm 1.5cm 0.0cm 2.5cm]{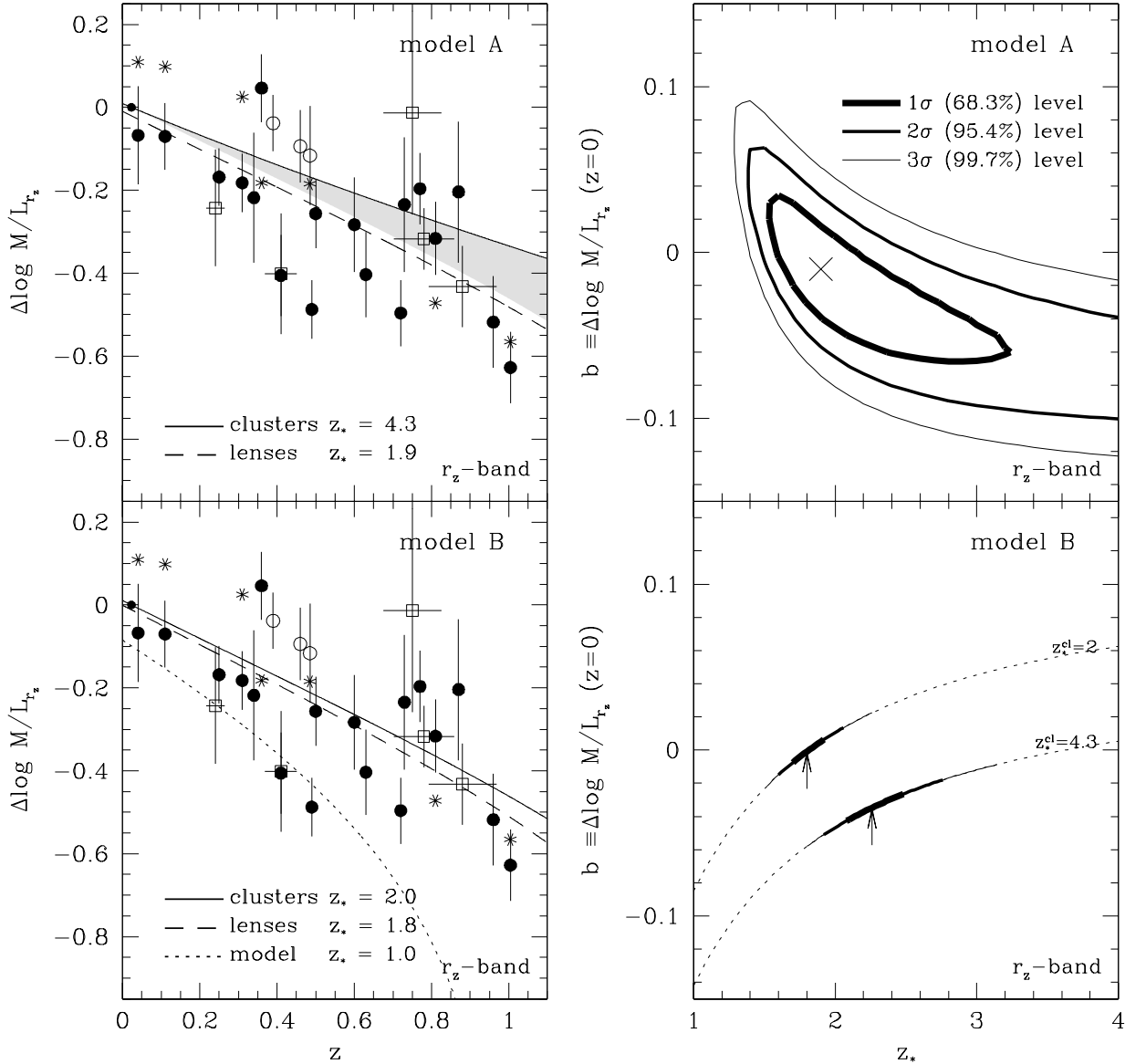}
  \caption{\slshape 
    \textbf{Upper-Left}:
    The evolution of the $M/L$ offset in the restframe Gunn $r$ band.
    The small dot is the average $M/L$ of the Coma galaxies.
    The lens galaxies are indicated with the same symbols as in
    Fig. \ref{fig:FPr}.  
    The solid line represents the single burst evolution of a stellar
    population formed at the mean cluster formation redshift of
    $z_\star=4.3$.  
    If we correct for progenitor bias, the latter formation redshift
    will be lower, and hence the evolution steeper.
    This is indicated by the shaded region, with $z_\star=2.0$ for
    maximum progenitor bias correction (vdF01).
    The dashed line is the best fit model for the lens galaxies, with
    $z_\star=1.9$ and zeropoint $b \equiv \Delta\log M/L_{r_z}(z=0)=-0.01$. 
    \textbf{Upper-Right}: 
    Confidence levels (1, 2 and 3$\sigma$) for combinations of $z_\star$
    and the zeropoint $b$ when they are both allowed to vary.
    \textbf{Lower-Left}: 
    The solid line shows the evolution for a stellar population formed
    at $z_\star=2.0$, which is the mean cluster value after maximum
    progenitor bias correction. 
    The dashed line is the model for a stellar population formed at the
    best-fitting star formation redshift of $z_\star=1.8$, if the lens
    and cluster galaxies are fitted in a self-consistent way (model B). 
    To illustrate the effect of changing $z_\star$ we also show a
    model with $z_\star=1$ (dotted line).
    \textbf{Lower-Right}: 
    Because cluster and lens galaxies are modeled self-consistently,
    the stellar formation redshift of the lens galaxies depends on
    that of the clusters $z_\star^{cl}$.
    The best-fit values vary from 1.8 to 2.3 (indicated by arrows) if
    $z_\star^{cl}$ is increased from 2.0 to 4.3, the values for
    maximum and no progenitor bias correction.
    The errors on the best-fit parameters from the confidence levels
    are multiplied with the square root of the reduced $\chi^2$ of the
    fit, to reflect the measured additional scatter compared to the
    expected scatter from the observational errors (see text and Table
    \ref{tab:z*lens} for resulting uncertainties on $b$ and $z_\star$).
    }
  \label{fig:MLr}
\end{figure*}

We assume that all early-type galaxies evolve in the same way, i.e. 
the coefficients $\alpha$ and $\beta$ are independent of redshift and
the same for cluster and field early-type galaxies.
Until now there is no convincing (observational) evidence against
these assumptions, but this might change if more and deeper data
become available (e.g.
van Dokkum \& Franx 1996\nocite{1996MNRAS.281..985V};
van Dokkum et al.\ 2001\nocite{2001ApJ...553L..39V};
Treu et al.\ 2002\nocite{2002ApJ...564L..13T}). 
If furthermore the effective radius and velocity dispersion do not
change with redshift, the difference in FP intercept is due to a
difference in surface brightness caused by luminosity evolution.
As a result, the evolution of the intercept of the FP is proportional
to the evolution of the mean $M/L$ ratio.
For each lens galaxy, the difference in FP intercept with respect to
the Coma FP (Fig. \ref{fig:FPr}) can be expressed as an offset in its
$M/L$ ratio relative to that of a galaxy of the same mass in Coma. 
Fig. \ref{fig:MLr} (two left panels) shows the dependence of
the $M/L$ offset on redshift, in the restframe $r$ band. 
The $M/L$ ratio of lens galaxies clearly shows a trend with redshift,
with the highest redshift objects having the lowest $M/L$ ratios.

From a linear fit to the $M/L$ ratios of the 26 lens galaxies, we find
an evolution rate $\d\log(M/L)/\d z$ of $-0.62 \pm 0.13$ and $-0.47
\pm 0.11$ in restframe $B$ and $r$ band respectively. 
The intercept of the fit, $\Delta\log M/L$ at $z=0$, is 
$-0.01 \pm 0.08$ and $-0.01 \pm 0.06$ respectively, so that locally
there is no significant offset in the mean $M/L$ ratio of the lens and
cluster galaxies. 
If this offset is forced to be zero, we find best-fit evolution rates
for the $B$ and $r$ band of $-0.63 \pm 0.06$ and $-0.49 \pm 0.05$,
which are only slightly higher than for the unconstrained fit.
(The $1\sigma$ errors from the fit are smaller than for the
unconstrained fit since only the slope is a fitting parameter, while
the intercept is fixed to zero.)
Restricting the fit to only the 21 lens galaxies with spectroscopic
redshifts, has no significant effect on the results. 
Similarly, we find no significant differences if we exclude the lens
galaxies from the fit for which a modest extrapolation was needed in
the transformation to restframe band; or if for the 7 lens galaxies
for which the velocity dispersion $\sigma_{c\star}$ measured from
stellar kinematics is available, this value is used instead of the
modeled $\sigma_c$ from the lensing geometry.

For cluster galaxies van Dokkum et al.\
(1998\nocite{1998ApJ...504L..17V}) find in the $B$ band an evolution
rate $\d\log(M/L)/\d z=-0.49 \pm 0.05$
This means on average a faster evolution for the lens galaxies, but
the difference is only significant if the $M/L$ ratio of lens galaxies
is forced to coincide with that of cluster galaxies at $z=0$.
The value for the cluster evolution rate by van Dokkum et al.\ 
(1998\nocite{1998ApJ...504L..17V}) does not take into
account possible effects due to morphological evolution. 
If early-type galaxies were transformed from late-type galaxies
at modest redshifts (e.g. Dressler et al.\
1997\nocite{1997ApJ...313...42D}) the early-type galaxies that were
already present at high redshift are only a subset of all progenitors
of low redshift early-type galaxies. 
As a result of this ``progenitor bias'' (vDF01) the formation redshift
of morphologically selected cluster early-type galaxies may be
overestimated. 
The lens galaxy sample is probably much less affected by progenitor
bias, since they are selected on mass and not on morphology, but due
to the merging of galaxies, a fraction of the progenitors may still
not be accounted for.  
After applying the maximum correction for progenitor bias allowed
by the data, vDF01 find a cluster evolution rate of 
$\d\log(M/L)/\d z=-0.56 \pm 0.05$ in the $B$ band, similar to the
evolution rate for the lens galaxies.  
These results suggest that, if there is no progenitor bias, field
galaxies may be younger than cluster galaxies, but that such an age
difference becomes less significant if we correct for possible
progenitor bias. 

Our results are consistent with determinations for field galaxies
based on direct spectroscopic measurements of the velocity
dispersions.   
For the $B$ band, 
van Dokkum et al.\ (2001\nocite{2001ApJ...553L..39V}) arrive at 
an evolution rate $\d\log(M/L)/\d z$ of $-0.59 \pm 0.15$, and  
Treu et al.\ (2002\nocite{2002ApJ...564L..13T}) find a value of
$-0.72_{-0.16}^{+0.11}$. 
Comparing our results for the $B$ band with those of R03, we find that
the evolution rate obtained by R03 is on average slower, but the
($1\sigma$) confidence limits still overlap.
For intercepts that are allowed to vary and forced to be zero, R03
find an evolution rate of $-0.54 \pm 0.09$ and $-0.56 \pm 0.04$
respectively. 

The error analysis and the transformation to restframe of R03
differs from ours (see also Section \ref{sec:introduction}), but we
cannot further investigate the effects of these differences, as R03 do
not give their resulting fundamental plane parameters. 
However, if we add (in quadrature) an constant additional error to the
observed uncertainties in the $M/L$ ratios of the lens galaxies, the
slope of the fit does decrease. 
Around the linear fit we measure a (biweight) scatter of $0.17$ for
the $B$ band and $0.15$ for the $r$ band\footnote{
The scatter in the $B$ band is probably higher than in the $r$ band
due to the fact that the $B$ band is more sensitive to recent star
formation.}, which in both cases is higher than the expected scatter
from the observational errors of $0.12$. 
If we now take for the constant additional scatter the difference (in
quadrature) between the measured and expected scatter, we find for the
unconstrained  fit in the $B$ band an evolution rate of $-0.56 \pm
0.12$, almost identical to the result of R03. 
If the offset is forced to zero the slope changes less, giving 
an evolution rate of $-0.61 \pm 0.05$.
Note that R03 rescale their input errors so that the best-fit model
has a reduced $\chi^2$ of one.
They assume that the additional scatter is due to underestimated
errors in the data set, whereas we propose that internal population 
differences cause the additional scatter.
Hence, we multiply our results with the square root of the reduced
$\chi^2$ to reflect this aspect, but we do not change the
uncertainties on the input data. 

The additional scatter implies that the $M/L$ ratios of the lens
galaxies are not well fitted with a single evolution rate, whereas for
the cluster galaxies the fit is very good (e.g. van Dokkum et al.\ 
1998\nocite{1998ApJ...504L..17V}).
This may be due to a significant spread in stellar population ages
among field galaxies, which induces an additional scatter measured in
the evolution rate $\d\log(M/L)/\d z$ of the lens galaxies
In the next Section, we relate the $M/L$ evolution to stellar
population ages by fitting simple single burst models.
In Section \ref{sec:colors}, we then study the restframe
colors of lens galaxies and their evolution. 
To find out whether there is a significant age spread, we also
investigate scatter in color and if it is correlated with
the scatter in $M/L$ ratio.  

\section{Stellar population ages}
\label{sec:ages}

The evolution of the $M/L$ ratio depends on the age of the
stellar population.
A stellar population which formed at low redshift will evolve faster
than a population formed at high redshift:
the luminosity of a young population becomes rapidly fainter when
short-lived massive and bright stars disappear, whereas the dimming
of the light is more gradual for an old population dominated by low
mass stars.
We estimate the stellar population ages of lens galaxies and the age
difference between lens and cluster galaxies by fitting simple single
burst models to the $M/L$ evolution.

The $M/L$ ratio of a single burst stellar population with
fixed mass evolves as
\begin{equation}
  \label{eq:singburstmod}
  M/L \propto (t - t_\star)^{\kappa},
\end{equation}
with $t_\star$ the stellar formation time, corresponding to a
stellar formation redshift $z_\star$
(e.g. Tinsley \& Gunn 1976\nocite{1976ApJ...206..525T}).
The coefficient $\kappa$ depends on the Initial Mass Function (IMF)
and metallicity, and also on the passband in which the luminosity is
measured. 
For a normal IMF with Salpeter (1955\nocite{1955ApJ...121..161S})
slope and Solar metallicity, $\kappa_B\approx 0.93$ and
$\kappa_r\approx 0.78$ for the restframe $B$ and $r$ bands
(Worthey 1994\nocite{1994ApJS...95..107W}).
Note that the predicted evolution is independent of $H_0$ because
the age dependence of the $M/L$ ratio is a power-law.

We first investigate if we can fit the observed $M/L$ evolution of the
lens galaxies by a single burst evolution model with the same stellar
formation redshift $z_\star$ as has been derived previously for
cluster galaxies (the null-hypothesis). 
In Section 5.2 and 5.3 we investigate the range of $z_\star$ allowed
by the data.

\subsection{Can cluster galaxies and lens galaxies have the same age?}
\label{sec:compcluster}

Before investigating more complex models we consider the case that
lens galaxies and cluster galaxies have the same luminosity weighted
stellar age. 
To determine the mean stellar formation redshift of cluster early-type
galaxies
$\langle z_\star^{cl} \rangle$ we fit a single burst
model  \eqref{eq:singburstmod} to previously published restframe
$B$ band data for the clusters 
CL1358+62 (Kelson et al.\ 1997\nocite{1997ApJ...478L..13K}), 
CL0024+16 (van Dokkum \& Franx 1996\nocite{1996MNRAS.281..985V}), 
MS2053+03 (Kelson et al.\ 1997\nocite{1997ApJ...478L..13K}) and 
MS1054-03 (van Dokkum et al.\ 1998\nocite{1998ApJ...504L..17V}) 
at redshifts 0.33, 0.39, 0.58 and 0.83, respectively, normalized with
respect to the average $M/L_B$ ratio of the local Coma cluster
at $z=0.023$ (J\o rgensen et al.\ 1996\nocite{1996MNRAS.280..167J}).

To perform such a single burst fit, we minimize 
\begin{equation}
  \label{eq:chisquared}
  \chi^2 = \sum_{i=1}^n \left( 
    \frac{ M/L_{\mathrm{mod},i}-M/L_{\mathrm{obs},i} }
    { \sigma(M/L_{\mathrm{obs},i}) } 
  \right)^2,  
\end{equation}
with $n$ the number of $M/L$ observations used as constraints,
$M/L_{\mathrm{obs},i}$ the $i$th observation, $M/L_{\mathrm{mod},i}$
the corresponding single burst prediction and
$\sigma(M/L_{\mathrm{obs},i})$ the uncertainty or error in this
observation.  
To determine the confidence levels, we calculate the difference in
$\chi^2$ between a model and the overall minimum, 
$\Delta \chi^2 = \chi^2 - \chi_{\mathrm{min}}^2$, to which the usual
Gaussian confidence levels can be assigned 
(e.g. Press et al.\ 1999\nocite{Press99..numrecipies}).  

The resulting best-fit mean formation redshift
$\langle z_\star^{cl} \rangle = 4.3_{-1.2}^{+3.7}$.
This direct fit does not take into account possible progenitor bias. 
After applying the maximum correction for progenitor bias allowed
by the data, vDF01 find $\langle z_\star^{cl} \rangle =
2.0_{-0.2}^{+0.3}$ for cluster galaxies.
Therefore, we consider $2.0<z_\star^{cl}<4.3$ as a plausible range 
for the mean star formation epoch of cluster early-type galaxies.

We fitted single burst models with this range of formation redshifts
to the restframe $B$ and $r$ band data of the lens galaxies. 
The fits improve towards higher formation epoch, but even those with
maximum progenitor bias correction are rejected with nearly
$100\,\%$.
We can allow for an offset between $M/L$ ratios of lens galaxies and
those of cluster galaxies. 
We will describe this offset with the value of $\Delta\log M/L$
at $z=0$, which we denote by $b$.
For a stellar formation epoch ranging from $4.3$ to $2.0$ (no to
maximum progenitor bias correction) the offset $b$ varies from $-0.12$
to $-0.06$ for the $B$ band, and  
from $-0.08$ to $-0.03$ for the $r$ band. 
These models with a systematic offset fit the lens data better, but
are still rejected with $>99.9\,\%$ confidence.

\subsection{Model A: unconstrained fit}
\label{sec:modelA}

We assume that the $M/L$ ratios of field galaxies are independent of
those of cluster galaxies, i.e. a cluster galaxy of a given age can
have a very different $M/L$ ratio than a field galaxy of the same
age. 
This may be the case if, e.g. the metallicities of field and cluster
galaxies are different at a given mass.  
In addition to the stellar formation redshift $z_\star$, we also have
the normalization of the single burst model as free parameter.   
We describe this second parameter with $b$ 
(see Section \ref{sec:compcluster}).  

For a range of $z_\star$ and $b$ values, we fit single burst models
the observed $M/L$ values of the redshifted lens galaxies.  
For the $r$ band, the 1,2 and 3 $\sigma$ limits on $z_\star$ and $b$
are shown the in the upper-right panel of Fig. \ref{fig:MLr}, with the
minimum indicated by a cross.   
The best-fit values of $z_\star$ and $b$ are
$1.8_{-0.5}^{+1.4}$ and $-0.03\pm0.09$ for the $B$ band,
and $1.9_{-0.6}^{+1.9}$ and $-0.01\pm0.07$ for the $r$ band 
(Table \ref{tab:z*lens}). 
Note that the given uncertainties are $1\sigma$ errors unless noted. 
The $M/L_r$ evolution that corresponds to the best-fit values is shown
in the upper-left panel of Fig.\ref{fig:MLr} (dashed line).  
Since for both $B$ and $r$ band the model parameter $b$ is not
significantly different from zero, we obtain similar best-fit values
for $z_\star$ if we normalize the single burst model such that 
$b \equiv 0$, i.e. if we assume that locally the average $M/L$ ratio
of field and cluster galaxies is the same.

\subsection{Model B: simultaneous fit to lens and
  cluster galaxies}
\label{sec:modelB}

Here we assume that the stellar populations of field and cluster
galaxies evolve in the same way. 
The stellar populations of the lens galaxies may form at a different
redshift than those of the cluster galaxies, but galaxies of a given
age have identical $M/L$ ratios. 
For the formation redshift of the stars in cluster galaxies
we use the values that we obtained in Section \ref{sec:compcluster}:
$z_\star^{cl} = 4.3$ if not corrected for progenitor bias
and $z_\star^{cl}=2.0$ after correction for maximum progenitor bias.
In both cases we determine the constraints on $z_\star$ for the
lens galaxies. 
Note that $z_\star$ and $b$ are coupled in this model, because of the
constraint that lens galaxies with the same age as cluster galaxies
have identical $M/L$ ratios. 
For the $r$ band the resulting constraints on $z_\star$ (and hence $b$)
are shown in the lower-right panel of Fig. \ref{fig:MLr}, with the
best fit values indicated by arrows.
From maximum to no progenitor bias correction, we find
$z_\star$ ranging from 1.6 to 2.4 ($1\sigma$) for the $B$ band, 
and from 1.6 to 2.6 ($1\sigma$) for the $r$ band 
(Table \ref{tab:z*lens}). 
For the case of maximum progenitor bias the $M/L$ evolution that
follows from the best-fit value $z_\star=1.8$ in the $r$ band, is
shown in the lower-left panel of Fig. \ref{fig:MLr} (dashed line). 

\begin{table}
  \begin{center}
    \begin{tabular}{lll}
      \hline
      model & $B_z$ band & $r_z$ band \\
     \hline
     A & 1.8  (-0.5/+1.4) & 1.9  (-0.6/+1.9) \\
     B (with $z_\star^{cl}=2.0$) 
     & 1.7  (-0.1/+0.2) & 1.8  (-0.2/+0.2) \\
     B (with $z_\star^{cl}=4.3$) 
     & 2.1  (-0.2/+0.3) & 2.3  (-0.3/+0.3) \\
     \hline \hline
   \end{tabular}
   \caption[]{\slshape 
     The mean formation redshift $\langle z_\star \rangle$ of a single
     burst stellar population for lens galaxies (1$\sigma$
     errors). Models as in Fig. \ref{fig:MLr} (see text for further
     details). 
    }
   \label{tab:z*lens}
  \end{center}
\end{table}

\subsection{Summary of results}
\label{sec:summaryresults}

We have demonstrated that the $M/L$ evolution of lens galaxies cannot
be fitted with models that provide good fits to cluster galaxies.
The fit clearly improves if lens galaxies are allowed to be
systematically offset from cluster galaxies due to metallicity
differences or other systematic effects.  

If we consider both this offset and the stellar formation epoch of the
lens galaxies as free parameters, the best-fit single burst model in
the $B$ and $r$ band are consistent. 
The resulting offset is not significantly different from zero, and
although the best-fit stellar formation redshift implies for lens
galaxies on average a younger stellar population than for cluster
galaxies, the resulting ($1\sigma$) range of 
$1.3 < \langle z_\star \rangle < 3.8$ is not significantly different
from the formation epoch for cluster galaxies of 
$1.8 < \langle z_\star^{cl} \rangle < 8.0$ for the full range from
no to maximum progenitor bias correction.

If we impose the constraint that galaxies of the same age have the
same $M/L$ ratio irrespective of their environment, we find
stellar formation redshifts for the lens galaxies of 
$1.6 < \langle z_\star \rangle < 2.0$ if the stars in cluster galaxies
formed at $z_\star^{cl}=2.0$ (maximum progenitor bias), 
and $1.9 <\langle z_\star \rangle < 2.6$ if the stars in cluster
galaxies formed at $z_\star^{cl}=4.3$ (no progenitor bias). 
In the latter case the stellar populations of the lens galaxies are
significant younger (10--15$\,\%$ at the present epoch) than those of
the cluster galaxies.  

If the local $M/L$ offset between lens and cluster galaxies is
allowed to vary, R03 find for the stellar formation redshift of the
lens galaxies a ($1\sigma$) range of 
$2.0 < \langle z_\star \rangle < 3.6$ from their $B$ band analysis.
Although we find a range of 
$1.3 < \langle z_\star \rangle < 3.2$ (model A, $B$ band) 
which implies on average a somewhat younger stellar population, the
results are consistent. 
R03 conclude that the $M/L$ evolution rates they measure favor old
stellar populations for the lens galaxies with a mean formation
redshift $\langle z_\star \rangle > 1.8$ at a $2\sigma$ confidence
level.
We find a lower $2\sigma$ confidence limit of 
$\langle z_\star \rangle > 1.2$. 

To test the hypothesis whether the $M/L$ evolution of the lens
galaxies can be fitted with a single burst model, we calculate the
reduced $\chi^2$.  
We also compare the measured scatter around the fit with the expected
scatter from the uncertainties in the $M/L$ ratios of the lens
galaxies.   
If we allow more freedom in the single burst models, the fit improves
(lower reduced $\chi^2$) and the measured scatter decreases.
However, even the over-all best-fit single burst model is rejected
with $>99\,\%$ confidence and the measured scatter of $0.17$ and
$0.15$ in $B$ and $r$ band is significant higher than the expected
scatter of $0.12$ from the observational errors. 
To establish whether the additional scatter is due to a
significant spread in ages among field galaxies, we study in the next
Section the (restframe) colors of the lens galaxies, and investigate
if there is also a significant scatter in the color and if it is
correlated with the scatter in their $M/L$ ratios.


\section{Colors}
\label{sec:colors}

If the stellar populations of lens galaxies are on average
younger than those of cluster galaxies (Section \ref{sec:ages}) we
expect their colors to evolve more rapidly and to be on average bluer
than those of cluster galaxies.   
For single burst stellar populations the $B-r$ color
evolves as
\begin{equation}
  \label{eq:singlburstcolevo}
  B-r = 2.5(\kappa_B-\kappa_r)\log(t-t_\star) + c,
\end{equation}
with $\kappa_B-\kappa_r \approx 0.15$ (Worthey
1994\nocite{1994ApJS...95..107W}) and $c$ a normalization constant.  

\begin{figure*}
  \begin{minipage}[t]{8.6cm}
    \begin{center}
      \includegraphics[draft=false,scale=0.45,trim=0cm 0.5cm 0cm 2.5cm]{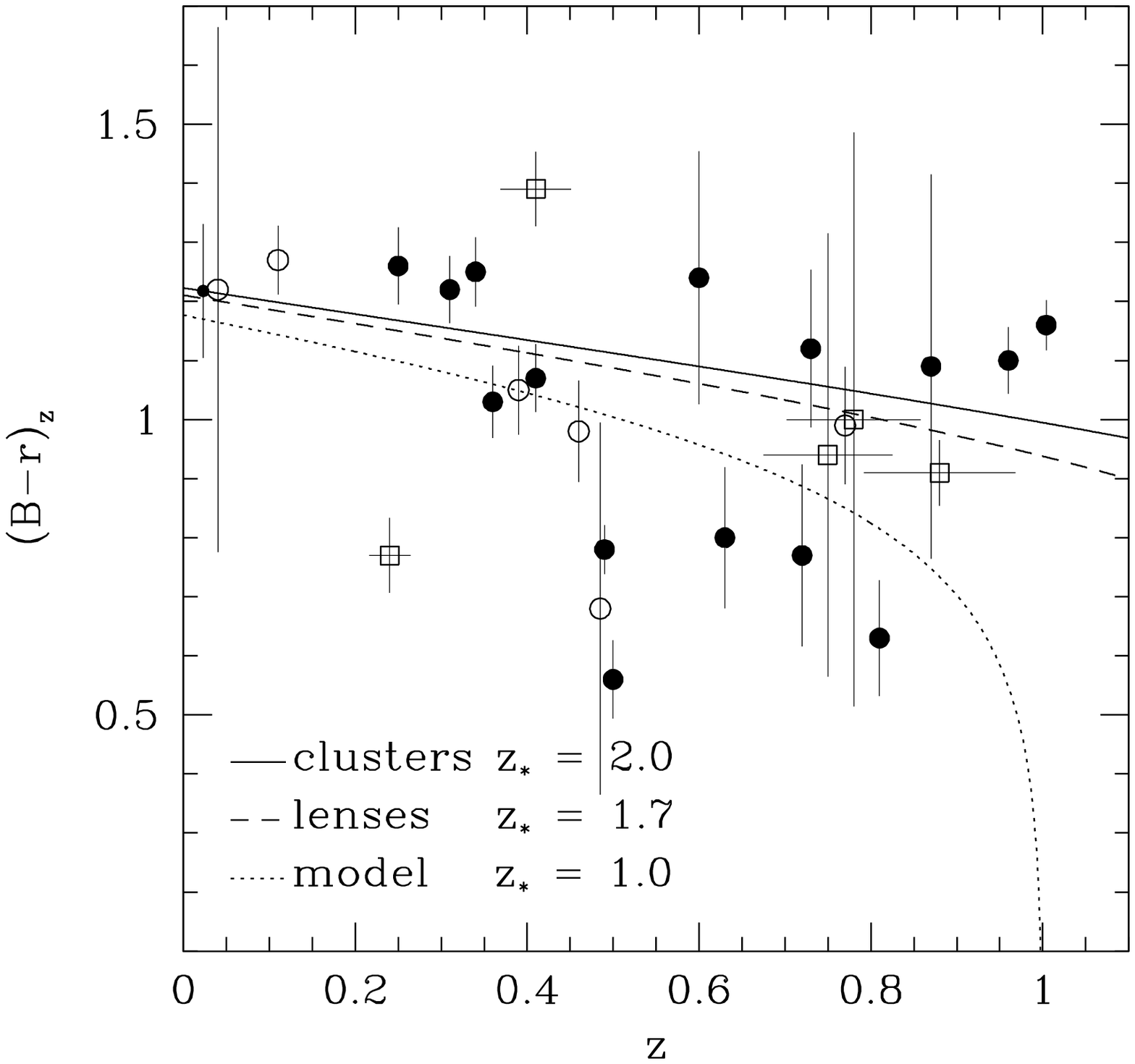} 
    \end{center}
  \end{minipage}
  \hfill
  \begin{minipage}[t]{8.6cm}
    \begin{center}
       \includegraphics[draft=false,scale=0.45,trim=0cm 0.5cm 0cm 2.5cm]{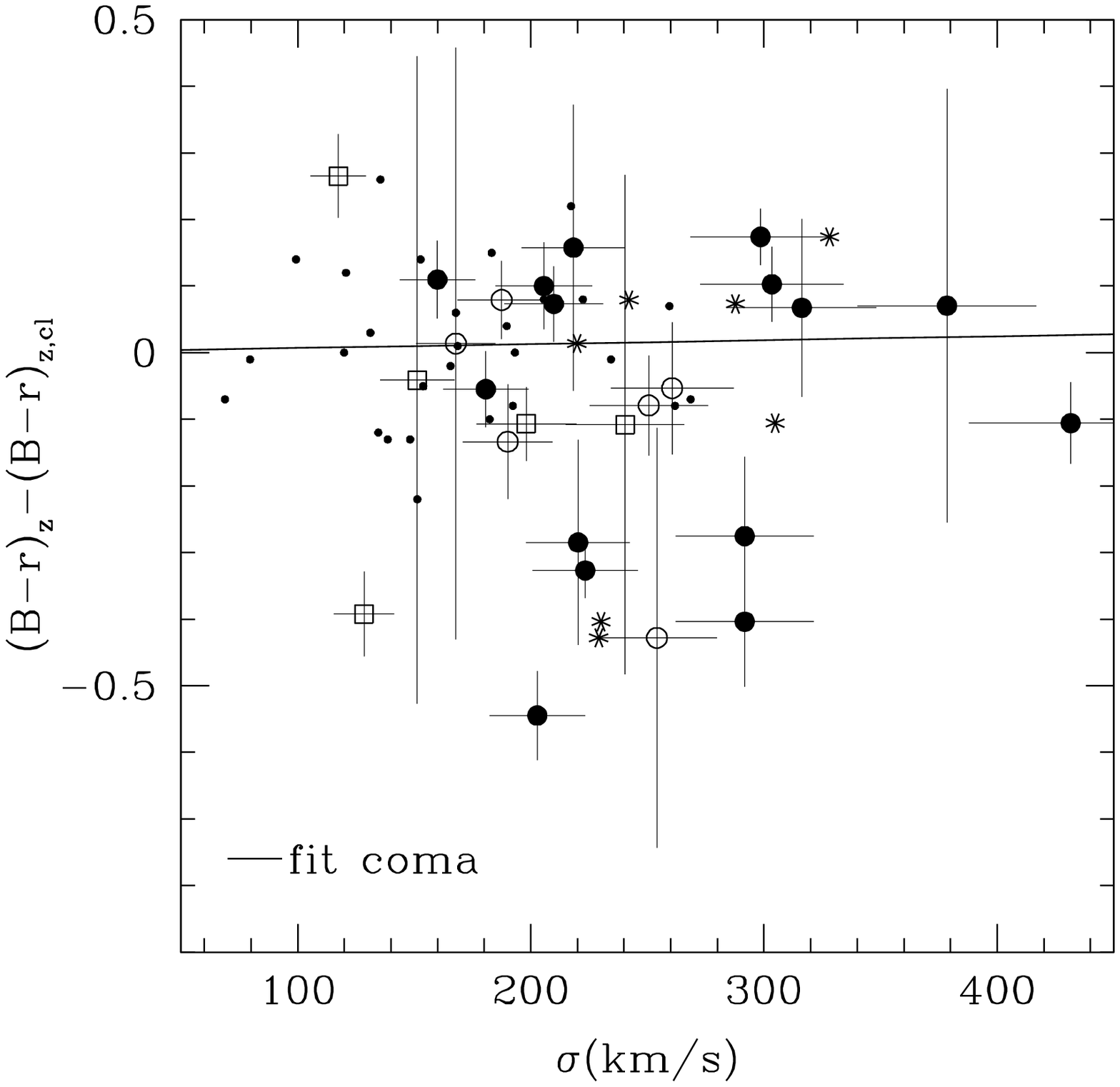} 
     \end{center}
   \end{minipage}
   \caption[]{\slshape 
    \textbf{Left:}
    Evolution of the restframe $B-r$ color.
    The symbols correspond to those in Fig. \ref{fig:FPr}, with the
    colors of the coma cluster galaxies averaged (small dot).
    The solid line shows the evolution for a stellar population formed
    at $z_\star=2.0$, which is the mean cluster value after maximum
    progenitor bias correction. 
    The dashed line is the model for a stellar population formed at the
    best-fitting star formation redshift of $z_\star=1.7$, if the lens
    and cluster galaxies are fitted in a self-consistent way. 
    To illustrate the effect of changing $z_\star$ we also show a
    model with $z_\star=1$ (dotted line).
    For redshifts beyond $z \sim 0.5$ the scatter in color increases
    and becomes significant higher than expected from the
    observational errors.    
    \textbf{Right:}
    The restframe $B-r$ colors of lens galaxies, after subtracting
    the fiducial model for cluster galaxies with $z_\star^{cl}=2.0$, 
    plotted versus velocity dispersion.  
    The solid line is a linear fit to the Coma cluster galaxies.
    On average the colors of the lens galaxies are bluer than those of
    the cluster galaxies, consistent with a younger average stellar
    population.  
    The scatter in the reduced colors is larger than expected from the
    observational errors. 
  }   
  \label{fig:color}
\end{figure*}

The left panel of Fig. \ref{fig:color} shows the restframe $B-r$
colors of the lens galaxies versus redshift.
For the lens galaxies with open circles, a modest extrapolation was
required in the transformation to either the restframe $B$ or $r$
band.  
The red outlier (at $z \sim 0.4$) is the lens BRI0952-0115, for
which the observed $R-H$ (F675W-F160W) color is significantly redder
than the modeled E/S0 color, and for which we had to extrapolate in
the transformation to restframe $B$ band. 
The single burst evolution of a stellar population formed at the
mean cluster formation epoch of $z_\star^{cl}=2.0$ (maximum
progenitor bias), is drawn (solid line) through the averaged color of
the coma cluster galaxies (small dot).
A decline in the colors of the lens galaxies with increasing redshift
is visible, in spite of the large scatter. 
If we fit single burst models as in Section \ref{sec:ages} (model A
and B), we find on average a younger stellar formation epoch for the
lens galaxies, but the difference with the cluster galaxies is never
significant.   
In case the stellar populations of field and cluster galaxies are
assumed to evolve in the same way (model B), the evolution model with
the best-fit stellar formation redshift of $z_\star=1.7$ is shown in 
the left panel of Fig. \ref{fig:color} (dashed line). 
Similarly as for the single burst model fits to the $M/L$ evolution,
the fits to the color evolution are rejected with $>99\,\%$
confidence, and the measured scatter is in all cases significantly
higher than the expected scatter from the observational errors in the
restframe colors of the lens galaxies.

To further investigate the scatter in color, we subtract the predicted
colors of cluster galaxies with $z_\star^{cl} = 2.0$, from the
restframe colors of the lens galaxies.     
Note that this choice of formation redshift corresponds to the 
\emph{minimum} age difference between field and cluster galaxies. 
In the right panel of Fig. \ref{fig:color} we show the resulting
residual colors plotted against velocity dispersion, with a linear fit
to the Coma cluster galaxies.  
The lens galaxies show a large spread in their reduced colors, and
are on average bluer by $\sim 0.1$ mag than the cluster galaxies.  
The bluer average color is qualitatively consistent with the on
average younger ages of lens galaxies derived from the single burst
model fits above and from the fits to the $M/L$ evolution 
(Section \ref{sec:summaryresults}). 
For the total sample of 26 lens galaxies, we measure a scatter in the
residual colors of $0.22$, that is significant higher than the
expected scatter of $0.18$.
From the left panel of Fig. \ref{fig:color}, we observe that the color
scatter increases beyond $z \sim 0.5$. 
For the 15 lens galaxies with $z \gtrsim 0.5$ we measure a scatter of
$0.24$, whereas for the lens galaxies with $z \lesssim 0.5$ the 
measured scatter is only $0.16$ and nearly identical to the expected
scatter of $0.15$ from the errors in the colors of these galaxies. 

\begin{figure}
  \begin{center}
    \includegraphics[draft=false,scale=0.45,trim=0cm 1.5cm 0cm 2.5cm]{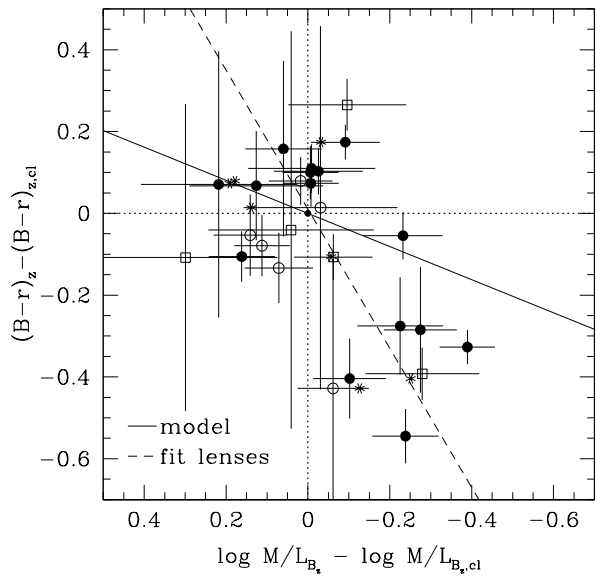} 
  \end{center}
  \caption[]{\slshape 
    The $B-r$ color versus $M/L_r$ ratio, after subtracting the
    fiducial cluster galaxy evolution model with $z_\star^{cl}=2.0$. 
    The symbols are the same as in Fig. \ref{fig:FPr}, with the
    average of Coma situated in the origin.
    The dashed line is a linear fit to the data; the solid line shows
    the expected correlation between color and $M/L$ ratio due to
    age variations, with slope 
    $2.5(\kappa_B-\kappa_r)/\kappa_B \approx 0.40$.
    Although the slope is uncertain, the correlation between color
    and $M/L$ ratio is significant at the $95\,\%$ level.
    Hence we interpret the intrinsic scatter in the colors of the
    lens galaxies as a stellar population effect, probably caused
    by a spread in their ages.
  }   
  \label{fig:colvsml}
\end{figure}

The additional measured scatter in the residual colors of the lens
galaxies may indicate a significant spread in the ages of the stellar
populations of the lens galaxies. 
We test whether the color spread is caused by a spread in
ages or other effects, by investigating whether the residual colors
correlate with the residual $M/L$ ratios. 
In Fig. \ref{fig:colvsml} we plot the color of the lens galaxies
against the logarithm of their $M/L$ ratio. 
For both quantities the expected evolution of a stellar population
formed at $z_\star^{cl} = 2.0$ was subtracted; hence a galaxy with the
average color and $M/L$ ratio of galaxies in Coma would 
be located at the origin. 
From Eq.\ \eqref{eq:singlburstcolevo} and \eqref{eq:singburstmod} it
follows that a linear relation is expected in Fig.\ \ref{fig:colvsml},
with a (time-independent) slope of
$2.5(\kappa_B-\kappa_r)/\kappa_B \approx 0.40$ (solid line). 
The data appear to be correlated in the expected sense, albeit
with large scatter.
To test whether the correlation is significant we used the Spearman's
rank-order correlation coefficient $r_S$.
We find that $r_S=0.47$, so that with $N=26$ lenses, we can reject
the hypothesis that the two quantities are uncorrelated
with $>95\,\%$ confidence.
The only viable explanation for the correlation is age variation. 
If the correlation would be caused by metallicity variations, (some)
field galaxies would be much less metal rich than cluster galaxies,
opposite to the result by 
Kuntschner et al. (2002\nocite{2002MNRAS.337..172K}).

The colors and $M/L$ ratios of the bluest lens galaxies are best
fitted with stellar formation redshifts as low as $z_\star \sim 1$. 
About half the lens galaxies are consistent with an old cluster-like
stellar population with stellar formation redshift 
$z_\star \gtrsim 2$.  
If galaxies form in a sequence of bursts, formation redshifts are
indicative of the last prominent epoch of star formation. 
The galaxies can have older underlying stellar populations.


\section{Summary and Conclusions}
\label{sec:discconcl}

We studied the evolution of the $M/L$ ratios of lens
galaxies in the restframe Johnson $B$ and Gunn $r$ bands.
For an flat cosmology with $\Omega_M=0.3$ and $\Omega_\Lambda=0.7$, 
we obtained an evolution rate $\d\log(M/L)/\d z$ of $-0.62 \pm 0.13$
in restframe $B$ and $-0.47 \pm 0.11$ in restframe $r$.
Due to differences in the determination of the FP parameters and the
corresponding errors, the evolution rate obtained by R03 of 
$-0.54 \pm 0.09$ is slightly slower but not significant different. 
Our results are consistent with determinations for field galaxies
based on direct spectroscopic measurements of the velocity dispersions.  
For the $B$ band, van Dokkum et al.\ (2001\nocite{2001ApJ...553L..39V}) 
arrive at an evolution rate of $-0.59 \pm 0.15$, and Treu et al.\
(2002\nocite{2002ApJ...564L..13T}) find a value of $-0.72_{-0.16}^{+0.11}$. 
The weighted mean of these results and our result yields an evolution
rate for field early-type galaxies of $-0.64 \pm 0.06$ in the $B$ band.  
For cluster galaxies 
van Dokkum et al.\ (1998\nocite{1998ApJ...504L..17V})
and vDF01 find an evolution rate between $\d\log(M/L_B)/\d z = -0.49
\pm 0.05$ and $-0.56 \pm 0.05$, for minimum and maximum progenitor
bias respectively.
The latter value is not significantly different
from the average evolution rate for the field galaxies.

We further investigated the $M/L$ evolution in Section \ref{sec:ages},
where we related it to stellar population ages by fitting simple
single burst models for a Salpeter (1955\nocite{1955ApJ...121..161S})
IMF and Solar metallicity. 
The $M/L$ evolution of cluster galaxies is well approximated by a
mean single burst formation redshift of 
$z_\star^{cl}=4.3_{-1.2}^{+3.7}$, and $z_\star^{cl}=2.0_{-0.2}^{+0.3}$
after maximum progenitor bias correction (vDF01).

We first tested if these cluster models could fit the $M/L$ evolution
of the lens galaxies.  
This is not the case, but the fits improve if there is a systematic
offset between lens and cluster galaxies of 
$\sim -0.1$ in $\Delta \log M/L$
Such an offset could be caused by, e.g. metallicity variations,
systematic differences in the velocity dispersions due to the
different measurement techniques, or other effects. 
It is interesting to note that hierarchical models have predicted such
a constant offset with redshift between cluster and field galaxies 
(see van Dokkum et al.\ 2001\nocite{2001ApJ...553L..39V}). 
However, similar to van Dokkum et al.\
(2001\nocite{2001ApJ...553L..39V}), we conclude that the observed
offset is much smaller than the predicted offset of 
$\Delta \log M/L_B \sim -0.26$. 

If we next allow the stellar formation redshift $z_\star$ also to
vary, we find for the best-fit single burst model that the offset is
not significant from zero, together with a mean stellar formation
epoch $\langle z_\star \rangle$ for the lens galaxies of 
$1.8_{-0.5}^{+1.4}$ in the $B$ band and
$1.9_{-0.6}^{+1.9}$ in the $r$ band. 
This means that on average the stellar populations of the lens
galaxies are younger, but the difference with the cluster galaxies is
not significant.
However, if we impose the constraint that galaxies of the same age
have the same $M/L$ ratio irrespective of their environment, we find
without correction for progenitor bias that the stellar populations of
the lens galaxies are significant younger (10--15$\,\%$ at the present
epoch) than those of the cluster galaxies.
In the case of maximum progenitor bias the average difference ($\sim
5\,\%$) is not anymore significant.

From their analysis R03 obtain a ($2\sigma$) lower limit for the mean
stellar formation epoch of lens galaxies of 
$\langle z_\star \rangle > 1.8$, whereas we find that lower stellar
formation redshifts are allowed, with $\langle z_\star \rangle > 1.2$
as a $2\sigma$ lower limit. 
Nevertheless, these results disagree with the prediction of semi-analytical
hierarchical galaxy formation models  
(e.g. Kauffmann 1996\nocite{1996MNRAS.281..487K};  
Kauffmann \& Charlot 1998\nocite{1998MNRAS.294..705K}; 
Diaferio et al.\ 2001\nocite{2001MNRAS.323..999D}) 
that early-type field galaxies in general have very late star formation
with $z_\star<1$.

Whereas the $M/L$ evolution of cluster galaxies is well approximated
by a single burst evolution model, we found that this is not the case
for the lens galaxies.
All single burst models are rejected with $>99\,\%$ based on the
reduced $\chi^2$, and the measured scatter is in all cases significant
higher than the expected scatter from the observational errors.
The additional scatter is most likely caused by differences in the
stellar population of the lens galaxies.

To investigate whether there is a significant spread in ages, we
studied in Section \ref{sec:colors} the restframe colors of the lens
galaxies and their evolution.
The colors of the lens galaxies are on average bluer than those of the
cluster galaxies, consistent with on a younger average stellar
population. 
We found that the measured scatter in the single burst fits to the
colors is much larger than the expected scatter from the  
observational errors.
Moreover, we showed that there is a significant correlation between
the colors and the $M/L_B$ ratios of the lens galaxies.
We interpret this as evidence for a significant spread in the stellar
population ages of the lens galaxies. 

Whereas about half of the lens galaxies are consistent with old
cluster-like stellar populations, the bluest galaxies are best fit by
single burst models with young stellar formation redshifts 
$z_\star \sim 1$.
For the seven blue lens galaxies with residual colors below $-0.2$ mag
in the right panel of Fig. \ref{fig:color}, we found (as a byproduct
of our transformation to restframe $B$ band) that only two of them are
best fitted by the E/S0 spectral type, i.e. $\sim 28\,\%$, whereas for
the total sample we found $77\,\%$. 
For the $r$ band even six of the seven blue galaxies are best fitted
by the Scd spectral type, instead of the ES/0 spectral type.  
Three of the seven blue galaxies indeed show (some) deviations from
early-type morphology.  
FBQ0951+2635 is an edge-on disk galaxy, SBS1520+530 is slightly
irregular and B1608+656 is an apparently dusty galaxy with
star forming regions.
However, a homogeneous sample of quasar subtracted (\texttt{NICMOS})
images of the lens galaxies is needed to do a more detailed and
systematic study of their morphologies. 
Such a study will be valuable to determine the cause of
apparently younger populations in a fraction of the lens galaxies.


\section*{acknowledgments}
\label{sec:acknowledgments}

We thank the anonymous referee for constructive and detailed comments,
which improved the paper significantly.


\bibliographystyle{mn2e}

\begin{thebibliography}{}

\bibitem{1990AJ....100...32B} 
  {Beers}, T.~C. and {Flynn}, K. and {Gebhardt}, K., 1990, AJ, 100, 32 

\bibitem{1998ApJ...493..529B}  
  {Bender} R.,  {Saglia} R.~P., {Ziegler} B.,  {Belloni} P.,
  {Greggio} L., {Hopp} U.,    {Bruzual} G.,  1998, 
  ApJ, 493, 529 

\bibitem{1990PASP..102.1181B} 
  {Bessell} M.~S.,  1990, PASP, 102, 1181

\bibitem{1987gady.book.....B}
  {Binney} J.,  {Tremaine} S.,  1987, {Galactic Dynamics}.
  Princeton, NJ, Princeton University Press

\bibitem{1989ApJ...345..245C} 
  {Cardelli} J.~A.,  {Clayton} G.~C., {Mathis} J.~S.,  1989, ApJ, 345, 245 

\bibitem{1980ApJS...43..393C} 
  {Coleman} G.~D.,  {Wu} C.~., {Weedman} D.~W.,  1980, ApJS, 43, 393 

\bibitem{2001MNRAS.323..999D}
  {Diaferio} A.,  {Kauffmann} G.,  {Balogh} M.~L.,  {White} S.~D.~M.,  {Schade}
  D.,    {Ellingson} E.,  2001, MNRAS, 323, 999

\bibitem{1987ApJ...313...59D}
  {Djorgovski} S.,  {Davis} M.,  1987, ApJ, 313, 59

\bibitem{1987ApJ...313...42D}
  {Dressler} A.,  {Lynden-Bell} D.,  {Burstein} D.,  {Davies} R.~L.,  {Faber}
  S.~M.,  {Terlevich} R.,    {Wegner} G.,  1987, ApJ, 313, 42

\bibitem{1997ApJ...313...42D}
  {Dressler} A., et al.\, 1997, ApJ, 490, 577

\bibitem{1987Faber}
  {Faber} S.~M.,  {Dressler} A.,  {Davies} R.~L.,  {Burstein} D.,  {Lynden-Bell}
  D.,  {Terlevich} R.,    {Wegner} G.,  1987, {Nearly Normal Galaxies}.
  Springer, New York, p. 175

\bibitem{1997ApJ...484...70F}
  {Falco} E.~E.,  {Shapiro} I.~I.,  {Moustakas} L.~A.,    {Davis} M.,  1997,
  ApJ, 484, 70

\bibitem{1992ApJ...386L..43F}
  {Foltz} C.~B.,  {Hewett} P.~C.,  {Webster} R.~L.,    {Lewis} G.~F.,  1992,
  ApJ, 386, L43

\bibitem{1994AJ....108.1476F}
  {Frei} Z.,  {Gunn} J.~E.,  1994, AJ, 108, 1476

\bibitem{1990ApJ...356..359H}
  {Hernquist} L., 1990, ApJ, 356, 359

\bibitem{2000...astroph9905116v4}
  {Hogg}, D.~W., 2000, astro-ph/9905116

\bibitem{2002ApJ...571..136I}
  {Im} M.,  {Simard} L.,  {Faber} S.~M.,  {Koo} D.~C.,  {Gebhardt} K.,  {Willmer}
  C.~N.~A.,  {Phillips} A.,  {Illingworth} G.,  {Vogt} N.~P.,    {Sarajedini}
  V.~L.,  2002, ApJ, 571, 136

\bibitem{1995MNRAS.273.1097J}
  {J\o rgensen} I.,  {Franx} M.,    {Kj\ae rgaard} P.,  1995a, MNRAS, 273, 1097

\bibitem{1995MNRAS.276.1341J}
  {J\o rgensen} I.,  {Franx} M.,    {Kj\ae rgaard} P.,  1995b, MNRAS, 276, 1341

\bibitem{1996MNRAS.280..167J}
  {J\o rgensen} I.,  {Franx} M.,    {Kj\ae rgaard} P.,  1996, MNRAS, 280, 167

\bibitem{1996MNRAS.281..487K}
  {Kauffmann} G.,  1996, MNRAS, 281, 487

\bibitem{1998MNRAS.294..705K}
  {Kauffmann} G.,  {Charlot} S.,  1998, MNRAS, 294, 705

\bibitem{1997ApJ...478L..13K}
  {Kelson} D.~D.,  {van Dokkum} P.~G.,  {Franx} M.,  {Illingworth} G.~D.,
  {Fabricant} D.,  1997, ApJl, 478, L13

\bibitem{1994ApJ...436...56K}
  {Kochanek} C.~S.,  1994, ApJ, 436, 56

\bibitem{1996ApJ...466..638K}
  {Kochanek} C.~S.,  1996, ApJ, 466, 638

\bibitem{2000ApJ...543..131K}
  {Kochanek} C.~S.,  {Falco} E.~E.,  {Impey} C.~D.,  {Leh{\'a}r} J.,  {McLeod}
  B.~A.,  {Rix} H.~.,  {Keeton} C.~R.,  {Mu{\~n}oz} J.~A.,    {Peng} C.~Y.,
  2000, ApJ, 543, 131 [K00]

\bibitem{2002ApJ...568L...5K}
  {Koopmans} L.~V.~E.,  {Treu} T.,  2002, ApJl, 568, L5

\bibitem{2003ApJ...583..606K}
  {Koopmans} L.~V.~E.,  {Treu} T.,  2003, ApJ, 583, 606

\bibitem{2002MNRAS.337..172K}
  {Kuntschner} H., {Smith} R.~J., {Colless} M., {Davies} R.~L.,
  {Kaldare} R., {Vazdekis}, A., 2002, MNRAS, 337, 172

\bibitem{1996AJ....111.1812L}
  {Leh{\' a}r} J., {Cooke} A.~J., {Lawrence} C.~R., {Silber} A.~D.,
  {Langston}, G.~I., 1996, AJ, 111, 1812

\bibitem{2000ApJ...536..584L}
  {Leh{\' a}r} J., et al.\, 2000, ApJ, 536, 584

\bibitem{1995ApJ...455..108L}
  {Lilly} S.~J.,  {Tresse} L.,  {Hammer} F.,  {Crampton} D.,    {Le Fevre} O.,
  1995, ApJ, 455, 108

\bibitem{1994ApJ...422..158O}
  {O'Donnell} J.~E.,  1994, ApJ, 422, 158

\bibitem{2002AJ....123.2903O}
  {Ohyama} Y., et al.\, 2002, AJ, 123, 2903

\bibitem{Press99..numrecipies}
  {Press} W.~H.,  {Teukolsky} S.~A.,  {Vettering} W.~T.,    {Flannery} B.~P.,
  1992, {Numerical Recipes}.
  Cambridge Univ. Press, Cambridge

\bibitem{1991Natur.350..211R}
  {Rhee} G.,  1991, Nature, 350, 211

\bibitem{2003ApJ...astroph0211229}
  {Rusin} D., {Kochanek}, C.~S., {Falco} E.~E., {Keeton} C.~R.,
  {McLeod} B.~A., Impey C.~D.,  {Leh{\' a}r} J., {Munoz} J.~A., 
  {Peng} C.~Y., {Rix} H.-W.,  2003, ApJ, accepted

\bibitem{1955ApJ...121..161S}
  {Salpeter} E.~E.,  1955, ApJ, 121, 161

\bibitem{1999ApJ...525...31S}
  {Schade} D.,  {Lilly} S.~J.,  {Crampton} D.,  {Ellis} R.~S.,  {Le F{\` e}vre}
  O.,  {Hammer} F.~.,  {Brinchmann} J.,  {Abraham} R.,  {Colless} M.,
  {Glazebrook} K.,  {Tresse} L.,    {Broadhurst} T.,  1999, ApJ, 525, 31

\bibitem{1998ApJ...500..525S}
  {Schlegel} D.~J.,  {Finkbeiner} D.~P.,    {Davis} M.,  1998, ApJ, 500, 525

\bibitem{1976PASP...88..543T}
  {Thuan} T.~X.,  {Gunn} J.~E.,  1976, PASP, 88, 543

\bibitem{1976ApJ...206..525T}
  {Tinsley} B.~M.,  {Gunn} J.~E.,  1976, ApJ, 206, 525

\bibitem{1998AJ....115....1T}
  {Tonry} J.~L.,  1998, AJ, 115, 1

\bibitem{1999ApJ...515..512T}
  {Tonry} J.~L.,  {Franx} M.,  1999, ApJ, 515, 512

\bibitem{2001MNRAS.326..237T}
  {Treu} T.,  {Stiavelli} M.,  {Bertin} G.,  {Casertano} S.,    {M{\o}ller} P.,
  2001, MNRAS, 326, 237

\bibitem{1999MNRAS.308.1037T}
  {Treu} T.,  {Stiavelli} M.,  {Casertano} S.,  {M\o ller} P.,    {Bertin} G.,
  1999, MNRAS, 308, 1037

\bibitem{2002ApJ...564L..13T} {Treu} T.,  {Stiavelli} M.,  {Casertano}
  S.,  {M\o ller} P.,    {Bertin} G., 2002, ApJl, 564, L13

\bibitem{1977A&A....54..661T} {Tully} R.~B.,  {Fisher} J.~R.,  1977, A\&A, 54, 661

\bibitem{1996MNRAS.281..985V}
  {van Dokkum} P.~G.,  {Franx} M.,  1996, MNRAS, 281, 985

\bibitem{2001ApJ...553...90V}
  {van Dokkum} P.~G.,  {Franx} M.,  2001, ApJ, 553, 90 [vDF01]

\bibitem{1998ApJ...504L..17V}
  {van Dokkum} P.~G.,  {Franx} M.,  {Kelson} D.~D.,    {Illingworth} G.~D.,
  1998, ApJl, 504, L17

\bibitem{2001ApJ...553L..39V}
  {van Dokkum} P.~G.,  {Franx} M.,  {Kelson} D.~D.,    {Illingworth} G.~D.,
  2001, ApJl, 553, L39

\bibitem{1994ApJS...95..107W}
  {Worthey} G.,  1994, ApJS, 95, 107

\end{thebibliography}


\bsp 

\label{lastpage}

\end{document}